\documentclass[conference]{IEEEtran}
\usepackage{cite}
\usepackage{amsmath,amssymb,amsfonts}
\usepackage{algorithmic}
\usepackage{graphicx,color}
\usepackage{textcomp}
\usepackage{epstopdf}
\usepackage{url}
\usepackage[protrusion=false]{microtype}

\def\BibTeX{{\rm B\kern-.05em{\sc i\kern-.025em b}\kern-.08em
    T\kern-.1667em\lower.7ex\hbox{E}\kern-.125emX}}
\begin{document}

\title{Rogue Drone Detection: A Machine Learning Approach}

\author{\IEEEauthorblockN{Henrik Ryd\'{e}n, Sakib Bin Redhwan, Xingqin Lin}
\text{Ericsson Research, Ericsson AB, Sweden}\\
Email: \{henrik.a.ryden, sakib.bin.redhwan,  xingqin.lin\}@ericsson.com}

\maketitle

\begin{abstract}
The emerging, practical and observed issue of how to detect rogue drones that carry terrestrial user equipment (UEs) on mobile networks is addressed in this paper. This issue has drawn much attention since the rogue drones may generate excessive interference to mobile networks and may not be allowed by regulations in some regions. In this paper, we propose a novel machine learning approach to identify the rogue drones in mobile networks based on radio measurements. We apply two classification machine learning models, Logistic Regression, and Decision Tree, using features from radio measurements to identify the rogue drones. We find that for high altitudes the proposed machine learning solutions can yield high rogue drone detection rate while not mis-classifying regular ground based UEs as rogue drone UEs. The detection accuracy however degrades at low altitudes.
\end{abstract}


\section{Introduction} \label{introduction}
Drones in general terms refer to any unmanned aerial vehicle. The uses of drones range from military applications to personal hobbies. Drone applications are creating broad socioeconomic benefits that are too large to ignore. According to the forecast of Goldman Sachs, a \$100 billion market opportunity for drones will emerge by 2020 \cite{rdd.gold17}. In this paper, we focus on civilian uses of drones. Besides recreational uses, example civilian drone use cases include precision agriculture, infrastructure monitoring and inspection, delivery, and photography \cite{rdd.gold17}, \cite{rdd.sesar16}. Technology should be in place to manage the growing fleet of drones \cite{rdd.frew08,rdd.ghar16,rdd.chand16,rdd.faa17,rdd.dot17}. Large enterprises are currently developing proprietary solutions suitable to their businesses, including proprietary communication technologies built on for example WiFi. Most of these connectivity solutions are short range and thus are not suitable for beyond visual line of sight drone operations. To extend the operation range, some drone manufacturers have started to support cellular connectivity in their products, such a product supporting Long-Term Evolution~(LTE) connection can be found in \cite{rdd.dji}. 
\newline \indent Mobile networks offer wide area, high speed, and secure wireless connectivity, which can significantly enhance control and safety of drone operations \cite{rdd.lin17}. Despite the great potential, there are challenges in using existing mobile networks optimized for ground usage to provide drone connectivity \cite{rdd.lin18}. The drones are typically flown at higher altitude than traditional ground user equipment~(UE). As the height above ground level increases, radio propagation becomes closer to line of sight free-space propagation. Though desirable for useful signal transmission, close to line of sight free-space propagation also implies that interfering signals can be strong if not properly managed \cite{rdd.yaj18}. The drones may be served by the side lobes of base station~(BS) antennas since the BS antennas are generally down-tilted to optimize terrestrial coverage. Many use cases require drones to transmit video feeds to their flight controllers, imposing heavy uplink traffic load on the networks. For better understanding of the performance of LTE for drones, we refer interested readers to the 3rd Generation Partnership Project~(3GPP) study item report on enhanced LTE support for aerial vehicles \cite{rdd.tr36777}. 
\newline \indent Due to the previously mentioned distinct features of drone UEs, it is important that the mobile networks can identify if a UE is a drone UE or a regular ground UE to provide the right service optimization for drone UEs while protecting the performance of ground UEs from the potential interfering signals from drone UEs. For legitimate drone UEs, standard mechanisms can be enforced so that these drone UEs can be recognized by the networks. For example, it can be required that a drone operator should acquire a Subscriber Identity Module (SIM) card that is designed or registered for drone use if the drone would like to use cellular connection. It is  very challenging to identify “rogue” drone UEs that are not registered with the networks as drones. This may occur when a normal ground UE is attached to a drone and being flown in the network. This phenomenon is being observed in the field and has drawn much attention from mobile operators, since flying a drone with regular UE may generate excessive interference to the network and flying a drone is not allowed by regulations in some regions. It is critical to identify these rogue drones from both mobile operators and security perspectives. Such a need has also been identified in the 3GPP study item on enhanced LTE support for aerial vehicles \cite{rdd.rp170779}.
\newline \indent In this paper, we propose a novel machine learning approach to identify rogue drones in the networks based on radio measurement reports sent by the UEs to the BSs. We apply two classification machine learning models, Logistic Regression, and Decision Tree, using features from radio measurements to identify the rogue drones. In Section \ref{mlApproach}, we describe in detail the proposed solutions and the evaluation methodology. In Section \ref{results}, we present the evaluation results for the proposed machine learning solutions and draw design insights from the results. We provide concluding remarks in Section \ref{conclusion}.

\section{A Machine Learning Approach} \label{mlApproach}
In this section, we describe the proposed machine learning approach to rogue drone detection in mobile networks.
\subsection{Simulation Scenario and Problem Formulation }

\begin{figure}[]
\centerline{\includegraphics[width=\columnwidth]{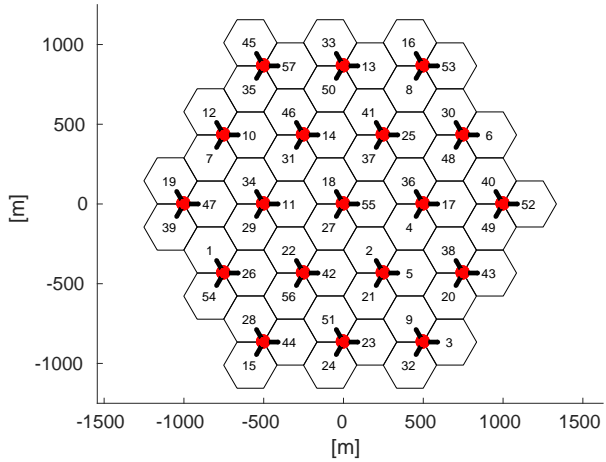}}
\caption{Simulation Deployment Scenario.}
\label{deployment}
\end{figure}

We study and evaluate our proposed machine learning approach to rogue drone detection under the agreed 3GPP scenarios \cite{rdd.tr36777}. Each simulation scenario includes a mix of outdoor drone UEs and regular ground UEs (that can be located indoor or outdoor). One of the basic deployment scenario is an Urban scenario with macro only homogeneous deployment. In this work, we utilize such a deployment with 19 sites and 3 sectors on each site, as shown in Figure \ref{deployment}. The inter-site distance is 500~m. Each BS has two cross-polarized antennas at the height of 25~m with a transmit power of 46~dBm. The carrier frequency is 2~GHz with 10~MHz bandwidth. The detailed channel model description can be found in \cite{rdd.tr36777}. In the simulation, 19,000 UEs were simulated for 60~seconds; 25~percent of them being drone UEs at different height above ground-level~(AGL) ranging from 15~m to 300~m; 65~percent being indoor ground UEs at different heights (modeling UEs located in high-rise buildings), and the rest of them being outdoor ground UEs at the height of 1.5~m. Although 25~percent drone UEs are likely to be over-dimensioned for a real network deployment, it was simulated to gather more data.

\begin{table}[]
\caption{Simulation parameters}
\begin{center}
\begin{tabular}{|p{2cm} p{2cm} p{3cm}|}
\hline
Simulation Parameter&\multicolumn{2}{p{5cm}|}{See \cite{rdd.tr36777} for detailed description} \\
\hline
& Sites& 19 sites in a hexagonal grid with 3 sectors per site \\
& BS height& 25 m \\
 & BS power& 46 dBm \\
Deployment& ISD& 500 m \\
& Carrier frequency& 2 GHz \\
& Carrier Bandwidth& 10 MHz \\
\hline
& Event& A3 \\
Mobility parameters& A3 offset& 2 dB \\
& A3 hysteresis& 1 dB \\
& TTT& 160 ms \\
\hline
&\multicolumn{2}{p{5cm}|}{Drone UEs: speed = \{120 km/h\}, height $\in$ \{15, 30, 60, 120, 300\} m}\\
UE properties&\multicolumn{2}{p{5cm}|}{Outdoor ground UEs: speed = \{120 km/h\}, height = \{1.5 m\}}\\
&\multicolumn{2}{p{5cm}|}{Indoor ground UEs:  speed = \{3 km/h\}, height $\in$  \{1.5, 11.5, 21.5, 31.5\} m}\\
\hline
UE classification&\multicolumn{2}{p{5cm}|}{19,000 UEs: 25\% drone UEs, 65\% indoor terrestrial UEs, 10\% outdoor terrestrial UEs}\\
\hline
\end{tabular}
\label{simParams}
\end{center}
\end{table}

\begin{figure}[]
\centerline{
\def\svgwidth{\columnwidth}
\begingroup%
  \makeatletter%
  \providecommand\color[2][]{%
    \errmessage{(Inkscape) Color is used for the text in Inkscape, but the package 'color.sty' is not loaded}%
    \renewcommand\color[2][]{}%
  }%
  \providecommand\transparent[1]{%
    \errmessage{(Inkscape) Transparency is used (non-zero) for the text in Inkscape, but the package 'transparent.sty' is not loaded}%
    \renewcommand\transparent[1]{}%
  }%
  \providecommand\rotatebox[2]{#2}%
  \newcommand*\fsize{\dimexpr\f@size pt\relax}%
  \newcommand*\lineheight[1]{\fontsize{\fsize}{#1\fsize}\selectfont}%
  \ifx\svgwidth\undefined%
    \setlength{\unitlength}{841.88976378bp}%
    \ifx\svgscale\undefined%
      \relax%
    \else%
      \setlength{\unitlength}{\unitlength * \real{\svgscale}}%
    \fi%
  \else%
    \setlength{\unitlength}{\svgwidth}%
  \fi%
  \global\let\svgwidth\undefined%
  \global\let\svgscale\undefined%
  \makeatother%
  \begin{picture}(1,0.70707071)%
    \lineheight{1}%
    \setlength\tabcolsep{0pt}%
    \put(0,0){\includegraphics[width=\unitlength]{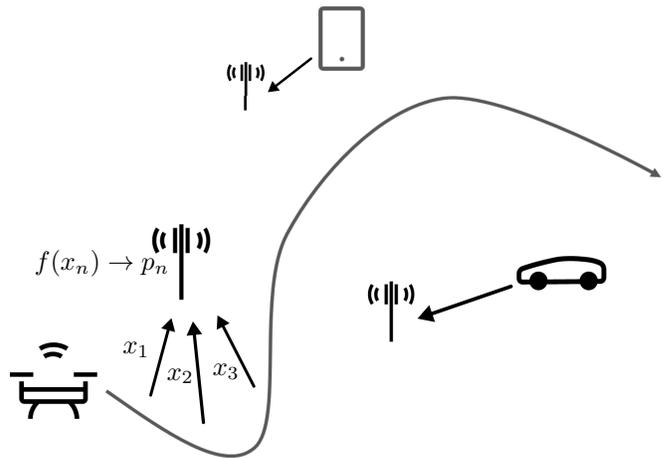}}%
    \put(0.3059651,0.12418579){\color[rgb]{0,0,0}\makebox(0,0)[lt]{\lineheight{1.25}\smash{\begin{tabular}[t]{l}$x_3$\end{tabular}}}}%
    \put(0.23644904,0.11811515){\color[rgb]{0,0,0}\makebox(0,0)[lt]{\lineheight{1.25}\smash{\begin{tabular}[t]{l}$x_2$\end{tabular}}}}%
    \put(0.16985665,0.15771063){\color[rgb]{0,0,0}\makebox(0,0)[lt]{\lineheight{1.25}\smash{\begin{tabular}[t]{l}$x_1$\end{tabular}}}}%
    \put(0.03667183,0.2845962){\color[rgb]{0,0,0}\makebox(0,0)[lt]{\lineheight{1.25}\smash{\begin{tabular}[t]{l}$f(x_n)\rightarrow p_n$\end{tabular}}}}%
  \end{picture}%
\endgroup%
}
\caption{A schematic overview of the technique: UEs transmit radio measurements $x_1,x_2,x_3$,…., over multiple time instances, and the network continuously predicts probability $p_n$ if a UE is a drone UE or not based on the measurement reports}
\label{solution}
\end{figure}

In the simulation, both drone UEs and outdoor terrestrial UEs have a speed of 120~km/h, while the indoor terrestrial UEs have a speed of~3 km/h. The main event that determines handover is so-called event A3 \cite{3gpp.36.331}, which is a handover measurement report triggering event when a neighbor cell becomes better than the serving cell. Event A3 is triggered if a neighbor cell measurement minus A3 hysteresis is greater than the serving cell measurement plus A3 offset. Once event A3 is triggered, the UE will wait for a predetermined time before it sends measurement reports to the serving cell. The predetermined time is called time-to-trigger~(TTT). Details of the simulation parameters can be found in Table \ref{simParams}. With the measurement reports transmitted by the UEs with 40 ms periodicity, the network continuously predicts if a UE is a drone UE or not. Figure \ref{solution} gives a schematic overview of this technique. 

The prediction can be mathematically expressed by 
\begin{equation}
f(x)\rightarrow p\label{eqModel}
\end{equation}

\noindent where $f$ is the per cell machine learning model, $x$ is the UE reported radio measurements, and $p$ is the probability of a UE being a drone UE. 
Since the simulated scenario is a homogeneous deployment, only one cell model is analyzed in this paper which is applicable for all cells in the scenario. For a heterogenous deployment, each cell may require its own model though the same principle applies. The proposed solution in this paper can be straightforwardly extended to heterogenous deployments.

\subsection{Training the Model}
Measurement data were collected from the previously mentioned LTE simulation. Collected data were divided into separate training and test sets. The model $f(\cdot)$ is trained with known legitimate drone UEs and regular ground UEs with the assumption that there are no rogue drone UEs in the training data set. Evaluation was performed with different data containing a mixture of drone UEs and regular ground UEs. 
The number of needed training samples depends on the scenario. In the homogenous scenario evaluated in this paper, as aforementioned only one model applicable for all cells is required. A more heterogenous scenario with one unique model per cell would substantially increase the number of needed training samples. Note that the training phase requires measurement data from known legitimate drone UEs flying in the network. The training phase can hence take long time in areas where drones are less likely to operate. Fortunately, rogue drones may likely be less of a problem in these areas as well.

\subsection{Machine Learning Model Selection}
Selection of machine learning model is a tradeoff between complexity and performance. A more complex model is likely to yield better performance if trained properly, but is also more resource consuming (time, memory, and computation resources). Depending on the problem at hand, a simple model may provide satisfactory performance while consuming much less resources \cite{rdd.kuhn13}. In this paper, we evaluate the following two basic classification machine learning models. Such an evaluation enables to decide on the needs for a complex model.

\textbf{Logistic Regression (LR):} studies the association between a categorical dependent variable and one or more independent variables. The posterior probability of the output class can be written as \cite{rdd.vai17} 

\begin{equation}
p=\frac{1}{(1+e^{-(\alpha+\beta_1 x_1+... ... +\beta_n x_n)})}\label{eqLRreg}
\end{equation}
\noindent where $\alpha$,$\beta_1$,\ldots,$\beta_n$  are the parameters of logistic regression and $x_1$,\ldots,$x_n$ are the features used for fitting into the model. In this paper, the output probabilities in the range [0,1] are categorized into two categories: 1 if it is a drone UE and 0~otherwise.

\begin{figure}[]
\centerline{\includegraphics[scale=0.5]{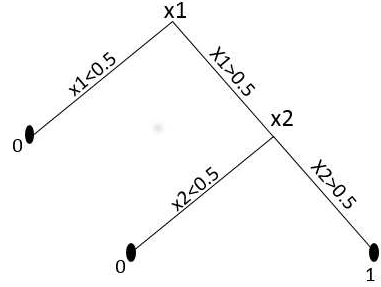}}
\caption{An example decision tree}
\label{dt}
\end{figure}

\textbf{Decision Tree (DT):} Decision trees, also known as classification trees or regression trees, are supervised learning methods used to create a model that predicts the value of a target variable by learning simple decision rules inferred from the data features. To predict a response, a leaf node is to be reached from the root node by following the decisions. In Figure \ref{dt}, an example DT based on two features is shown. For each feature, a decision is taken after a leaf node is reached.

\subsection{Feature Selection}
During post-processing, features for machine learning were extracted from the simulated data. In this paper, the following four features are considered:
\begin{itemize}
\item RSSI: Received signal strength indicator 
\item RSRP-STD: Standard deviation of the eight strongest reference signals received powers (RSRP)  
\item RSRP-gap: Difference between the strongest RSRP and the second strongest RSRP
\item Serving cell RSRP
\end{itemize}
Combinations of these features were used to evaluate the performance. However, it should be noted that the features and their combinations are not restricted to the above. 
The selection of these features is made based on the key observation that a drone UE operating at a high altitude is expected to have close to line of sight free-space propagation environment that leads to low variance of RSRPs of the strongest cells. RSSI statistics of drone UEs are different from those of regular ground UEs due to similar reason, e.g., a drone UE may receive signals from multiple cells with similar strengths.

\subsection{Performance Requirements and Metrics}
In machine learning classification, the false positive rate~(FPR), also known as false alarm rate, is the ratio of the number of incorrectly classified negatives (False Positives) to the total number of negatives (False Positives + True Negatives). In the context of this paper, FPR $>$ 0 means that some regular ground UEs are being identified as rogue drone UEs. Depending on how the networks deal with the UEs that are labeled as “rogue drone UEs,” the cost of false positives may be high since false positives may lead to unpleasant user experience. Therefore, one should try to achieve a very low FPR (ideally, zero FPR) during the classification. Aiming at zero FPR minimizes the number of occurrences that regular ground UEs are labeled as rogue drone UEs in our investigated test set of UEs. Therefore, the first metric studied in this paper is the drone detection rate (i.e., true positive rate) at zero FPR, i.e., the ratio of the number of correctly classified positives to the total number of positives while ensuring no regular ground UEs are labeled as drone UEs in our test set.  Note that this only guarantees that we have no ground UEs labeled as rogue drone in the test set of UEs. If a ground UE sends a measurement report different from the reports when building the model, there is a risk of classifying it as a rogue drone.
This paper also considers the ROC (Receiver Operating Characteristic) AUC (Area Under the Curve) classification performance metric \cite{rdd.han82}. The value of AUC lies in the range [0, 1]. The higher the AUC value, the better the prediction accuracy. For a model to be acceptable, AUC has to be greater than 0.5. An AUC of 1 represents a perfect classifier and the goal is to achieve an AUC as close to 1 as possible. 

\section{Results and Analysis}\label{results}
In this section, we first provide a comparison of the two trained machine learning models: LR and DT, followed by an example prediction of the three types of UEs: drone UEs, indoor ground UEs, outdoor ground UEs. We then evaluate the prediction performance using all test UEs in the simulated network, and observe how the accuracy depends on the drone altitude. 
\subsection{Model Training and Examples}
\begin{figure}[]
\centerline{\includegraphics[width=\columnwidth]{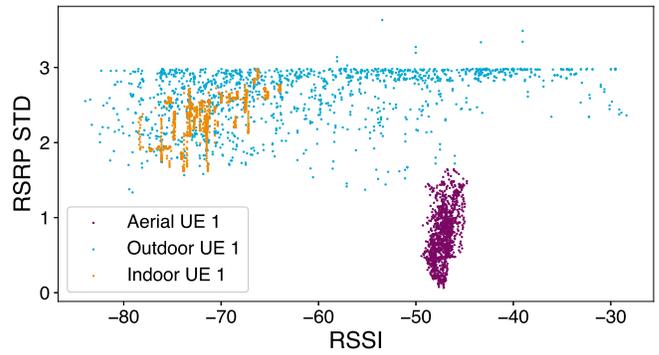}}
\caption{ \{RSSI, RSRP-STD\} samples for three UEs of different types }
\label{rssi_rsrp_for_threeUEs}
\end{figure}

The prediction accuracy is evaluated by first exploring the models using the two features: RSSI and RSRP-STD. The models are trained using the test UEs (50~percent of the simulated data). Next, we give an example prediction of three UEs of different types: a drone/aerial UE, an indoor ground UE, an outdoor ground UE. Figure \ref{rssi_rsrp_for_threeUEs} shows all received {RSSI, RSRP-STD} samples over 60 seconds in the simulation. It can be seen that the {RSSI, RSRP-STD} region for the indoor UE or the aerial UE is much smaller than that of the outdoor UE. The indoor UE is expected to have similar {RSSI, RSRP-STD} values since it is almost static. For the drone/aerial UE, despite moving at the high speed of 120 km/h, the RSRP-STD values are small and the RSSI values do not vary much during the simulation. This is because the radio environment in the sky is close to line of sight free-space propagation and consequently the signals from both serving cell and interfering cells decay much more slowly compared to the propagation on the ground.

\begin{figure}[]
\centerline{\includegraphics[width=\columnwidth]{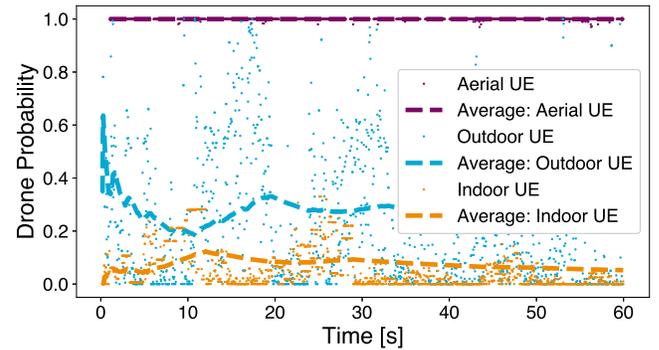}}
\caption{Drone UE detection probability versus simulation time}
\label{probability_over_time_three_UEs}
\end{figure}

After training the model, we first illustrate the prediction procedure for the three different UEs from Figure \ref{rssi_rsrp_for_threeUEs}. Figure \ref{probability_over_time_three_UEs} shows the drone UE detection probability over time when using DT classifier with RSSI and RSRP-STD as features. In the figure, we also plot the running average drone UE detection probability that is used for the final prediction, which corresponds to the average of the probability sequence $ p_1$ ,$p_2$ ,$p_3$ ,\ldots , $p_T$, where $T$ is bounded by the 60 seconds simulation time, and the time between samples that is 40ms. The figure shows that using the averaged prediction the probability that an outdoor ground UE is labeled as a drone UE decreases from approximately 0.6 after a few measurements to approximately 0.2 after 60 seconds.

\subsection{Prediction Performance}
\begin{figure}[]
\centerline{\includegraphics[scale=0.45]{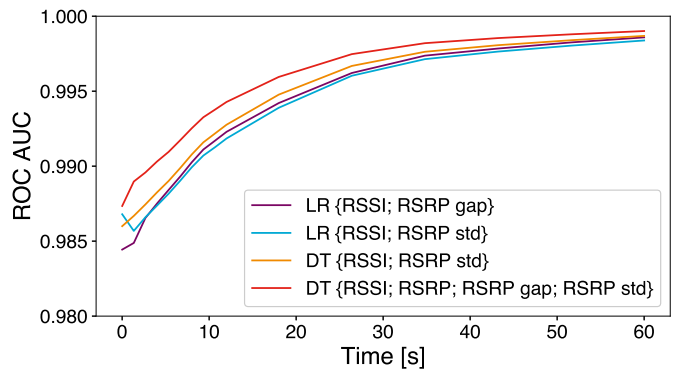}}
\caption{ROC AUC values versus simulation time}
\label{detected_drones_auc}
\end{figure}

\begin{figure}[ht]
\centerline{\includegraphics[width=\columnwidth]{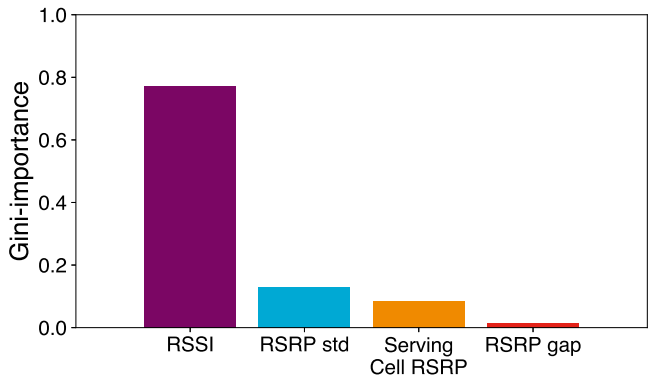}}
\caption{Feature importance for the DT classifier}
\label{feature_importance}
\end{figure}

\begin{figure}[]
\centerline{\includegraphics[width=\columnwidth]{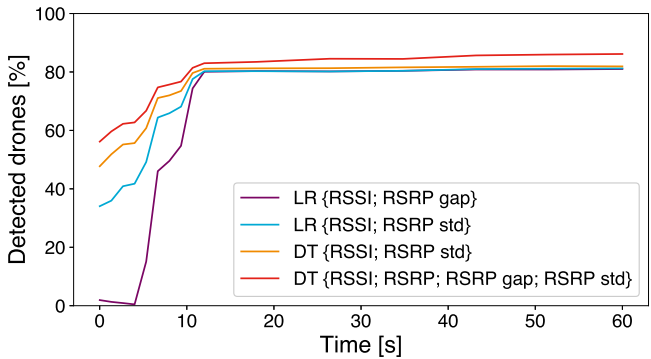}}
\caption{Drone detection rate with zero FPR in the UE test set}
\label{detected_drones_fpr0}
\end{figure}

\begin{figure}[]
\centerline{\includegraphics[scale=0.45]{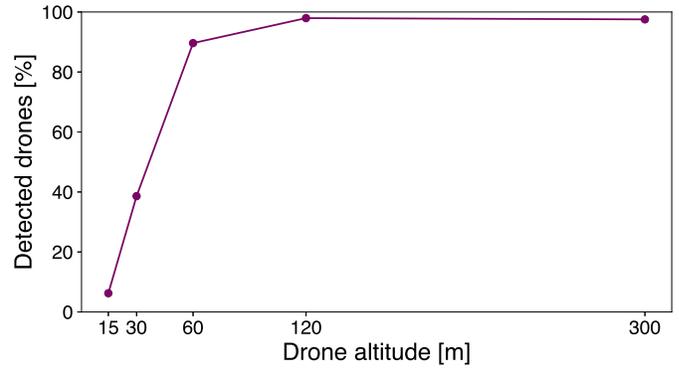}}
\caption{Detected drone per altitude after 60s using the DT classifier with features \{RSSI,RSRP-STD\}}
\label{correct_classified_height}
\end{figure}

As mentioned in the previous section, the ROC AUC metric is a good accuracy indicator of the model prediction performance. The ROC AUC curves presented in Figure~\ref{detected_drones_auc} show the prediction performance of both models (LR and DT) with several different combinations of the features. It can be seen from Figure~\ref{detected_drones_auc} that the DT classifier with all the four extracted features provides the best performance. However, the four features are not equally important in the DT classifier. 
We analyze the feature importance by using the Gini importance metric which is based on the Gini impurity in the trained DT model \cite{rdd.step09}. The feature importance is plotted in 
 Figure \ref{feature_importance}. As can be seen from the figure, the RSRP-gap feature has the least amount of impact on the result. This suggests that replacing the RSRP-gap feature with a more important feature would likely yield better performance, or removing the RSRP-gap feature can save resources while maintaining similar prediction accuracy. This is a tradeoff that deserves further consideration in practice. Figure \ref{detected_drones_fpr0} further shows the drone detection rate while maintaining zero FPR. It is found that using all the four features with the DT classifier yields the best performance. However, all the methods yield greater than 80~percent detection rates in the simulated scenario after 12~seconds, and the detection rates only improve minimally as more samples are added. During the first 10 seconds in the simulation, the DT classifiers show better performance than the LR classifiers. 
 To further analyze the performance, we separate the drone UE detection accuracy statistics based on the height above ground-level of the drone UEs. In Figure \ref{correct_classified_height}, we show the percentage of detected drone UEs at each height above ground-level after 60~seconds using the DT classifier with {RSSI, RSRP-STD} as the features. It can be seen from Figure \ref{correct_classified_height} that high drone UE detection rates can be achieved at the heights greater than 60~m. In contrast, only 5~percent of the drone UEs can be identified at 15~m while meeting the zero FPR target. The undetected drone UEs are at or below 60~m and they account for approximately 20~percent of all the drone UEs, leading to approximately 80~percent overall detection rate shown in Figure~\ref{detected_drones_fpr0}.
The observations in Figure \ref{correct_classified_height} may be explained as follows. For drone UEs above 60~m, their radio propagation environments are quite different from the radio propagation environment on the ground. As a result, high detection accuracy can be achieved for drone UEs above 60~m based on the radio measurements reported by the UEs. For drone UEs below 60 m, especially at the height of 15~m, the radio propagation environments of the drone UEs are similar to the radio propagation environment on the ground. As a result, it becomes challenging to distinguish these drone UEs from regular ground UEs solely based on the reported radio measurements.
\subsection{An Implementation Remark}
\begin{figure}[]
\centerline{
\def\svgwidth{220pt}
\begingroup%
  \makeatletter%
  \providecommand\color[2][]{%
    \errmessage{(Inkscape) Color is used for the text in Inkscape, but the package 'color.sty' is not loaded}%
    \renewcommand\color[2][]{}%
  }%
  \providecommand\transparent[1]{%
    \errmessage{(Inkscape) Transparency is used (non-zero) for the text in Inkscape, but the package 'transparent.sty' is not loaded}%
    \renewcommand\transparent[1]{}%
  }%
  \providecommand\rotatebox[2]{#2}%
  \newcommand*\fsize{\dimexpr\f@size pt\relax}%
  \newcommand*\lineheight[1]{\fontsize{\fsize}{#1\fsize}\selectfont}%
  \ifx\svgwidth\undefined%
    \setlength{\unitlength}{501.7500173bp}%
    \ifx\svgscale\undefined%
      \relax%
    \else%
      \setlength{\unitlength}{\unitlength * \real{\svgscale}}%
    \fi%
  \else%
    \setlength{\unitlength}{\svgwidth}%
  \fi%
  \global\let\svgwidth\undefined%
  \global\let\svgscale\undefined%
  \makeatother%
  \begin{picture}(1,0.74588937)%
    \lineheight{1}%
    \setlength\tabcolsep{0pt}%
    \put(0,0){\includegraphics[width=\unitlength]{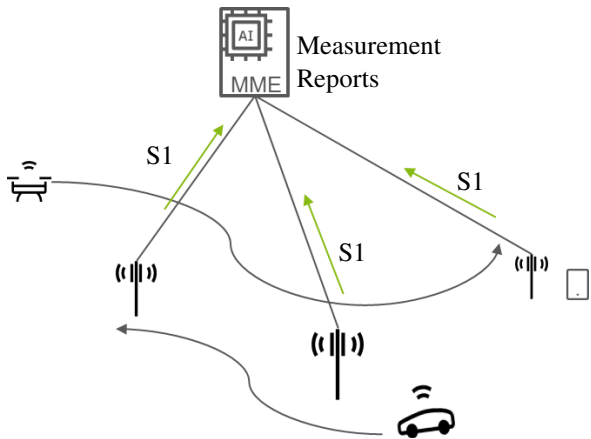}}%
    \put(0.7730087,0.42600897){\color[rgb]{0,0,0}\makebox(0,0)[lt]{\lineheight{1.25}\smash{\begin{tabular}[t]{l}S1\end{tabular}}}}%
    \put(0.56930072,0.3028428){\color[rgb]{0,0,0}\makebox(0,0)[lt]{\lineheight{1.25}\smash{\begin{tabular}[t]{l}S1\end{tabular}}}}%
    \put(0.23831628,0.47047037){\color[rgb]{0,0,0}\makebox(0,0)[lt]{\lineheight{1.25}\smash{\begin{tabular}[t]{l}S1\end{tabular}}}}%
    \put(0.49754427,0.65983342){\color[rgb]{0,0,0}\makebox(0,0)[lt]{\lineheight{1.25}\smash{\begin{tabular}[t]{l}Measurement\\Reports\end{tabular}}}}%
  \end{picture}%
\endgroup%
}
\caption{Network implementation: Cells sending reports to central entity, for example MME in LTE context}
\label{central_impl}
\end{figure}

\begin{figure}[]
\centerline{
\def\svgwidth{220pt}
\begingroup%
  \makeatletter%
  \providecommand\color[2][]{%
    \errmessage{(Inkscape) Color is used for the text in Inkscape, but the package 'color.sty' is not loaded}%
    \renewcommand\color[2][]{}%
  }%
  \providecommand\transparent[1]{%
    \errmessage{(Inkscape) Transparency is used (non-zero) for the text in Inkscape, but the package 'transparent.sty' is not loaded}%
    \renewcommand\transparent[1]{}%
  }%
  \providecommand\rotatebox[2]{#2}%
  \newcommand*\fsize{\dimexpr\f@size pt\relax}%
  \newcommand*\lineheight[1]{\fontsize{\fsize}{#1\fsize}\selectfont}%
  \ifx\svgwidth\undefined%
    \setlength{\unitlength}{1176.7500173bp}%
    \ifx\svgscale\undefined%
      \relax%
    \else%
      \setlength{\unitlength}{\unitlength * \real{\svgscale}}%
    \fi%
  \else%
    \setlength{\unitlength}{\svgwidth}%
  \fi%
  \global\let\svgwidth\undefined%
  \global\let\svgscale\undefined%
  \makeatother%
  \begin{picture}(1,0.68706179)%
    \lineheight{1}%
    \setlength\tabcolsep{0pt}%
    \put(0,0){\includegraphics[width=\unitlength]{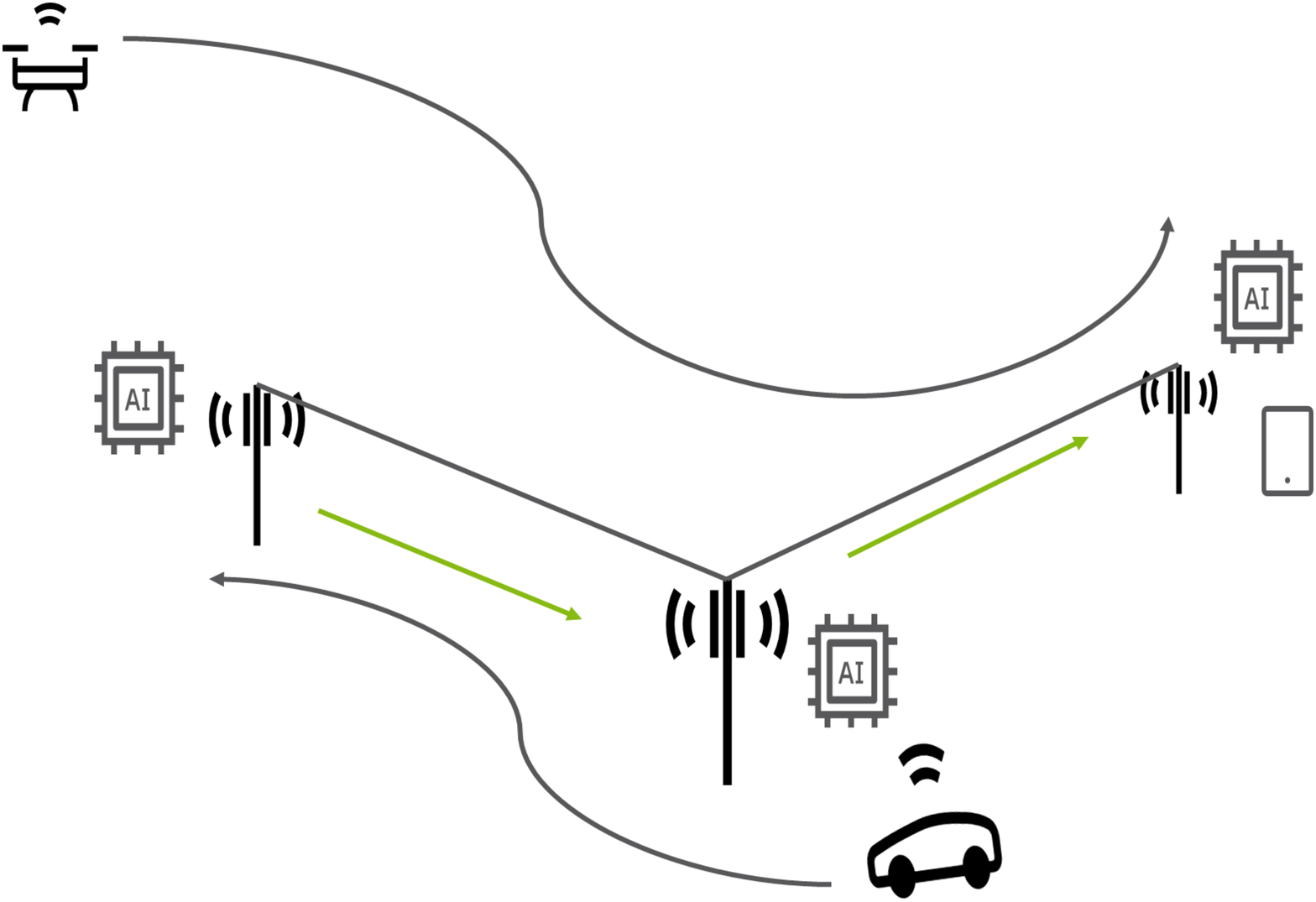}}%
    \put(0.30720512,0.29469014){\color[rgb]{0,0,0}\rotatebox{-22.384559}{\makebox(0,0)[lt]{\lineheight{1.25}\smash{\begin{tabular}[t]{l}X2\end{tabular}}}}}%
    \put(0.75020147,0.28007773){\color[rgb]{0,0,0}\makebox(0,0)[lt]{\lineheight{1.25}\smash{\begin{tabular}[t]{l}X2\end{tabular}}}}%
    \put(0.07649535,0.56705681){\color[rgb]{0,0,0}\rotatebox{-22.823429}{\makebox(0,0)[lt]{\lineheight{1.25}\smash{\begin{tabular}[t]{l}Measurement Reports\\ Prediction results\end{tabular}}}}}%
  \end{picture}%
\endgroup%
}

\caption{Network implementation: Cells sharing reports and results}
\label{distributed_impl}
\end{figure}

In the proposed solution, the rogue drone UE detection is based on a time-series of drone UE prediction probabilities, and the performance increases with the sequence length. This requires that in a real network the prediction probabilities from one cell are combined with the prediction probabilities from the next cell(s). One straightforward method is to forward the per cell sequence of prediction probabilities to a central entity, for example, the mobility management entity~(MME) as shown in Figure~\ref{central_impl}. The MME can then make the rogue drone UE detection based on the received probabilities from one or more cells. Another method is to exchange the measurements or the per cell sequence of prediction probabilities via X2 interface between cells as outlined in Figure~\ref{distributed_impl}. Each cell can run own algorithm(s) and forward the results with predicted values to the neighbor cell, the process can be terminated when a decision (a UE being drone/regular) is reached with certain precision.  

To avoid large computational and memory requirements on the network nodes, a two step procedure can be followed where the first step consists of classifying UEs into possible drones with a low probability of classifying drone UEs as a regular UE whilst having a high probability to classifying regular UEs as possible drone UEs. The first step can be a low complexity ML model or some pre-defined rules based on statistics. The second step can then refine the results on the subset of UEs identified in the first step by applying the methods described on this paper.

\section{Conclusion}\label{conclusion}
In this paper, we propose a novel machine learning approach to identify rogue drones in the mobile networks based on radio measurement reports. We have studied two classification machine learning models, Logistic Regression and Decision Tree, under different combinations of the four features: RSSI, RSRP-STD, RSRP-gap, and serving cell RSRP. We evaluate the solutions in a homogeneous network according to the 3GPP simulation assumptions. There is a tradeoff between rogue drone UE detection accuracy and false positives that classify regular ground UEs as rogue drone UEs. In the simulated scenario, we find that the proposed machine learning approach can achieve 100~percent detection rate for rogue drone UEs above 60~m height while meeting 0~percent false positive rate. The detection accuracy, however, degrades at lower heights: only 5~percent detection rate for rogue drone UEs at the height of 15~m to meet the 0~percent false positive rate. However, low altitude flying drones is less likely to create more interference than a regular UE and thus identification of them are less crucial from a network management point. Future work can consider more sophisticated deployments and/or different features to optimize the proposed machine learning approach. Also, the action after identifying a rogue drone should be investigated. One possible action after detecting a rogue drone UE is to warn the UE that it is flying without permission. The warning could possibly indicate a time when the connection will be terminated, enabling the drone to first safely return to ground level.

\bibliography{3gpp,rdd} 
\bibliographystyle{ieeetr}

\end{document}